\documentclass[preprint,superscriptaddress,aps,floatfix]{revtex4}
\usepackage[english]{babel}
\usepackage{graphicx}
\usepackage{amssymb,amsmath,amsfonts}
\usepackage[utf8]{inputenc}
\usepackage[T1]{fontenc}

\begin{document}

\title{Josephson junction microwave amplifier in self-organized noise compression mode}
\author{Pasi L\"ahteenm\"aki$^{*\dagger}$}
\affiliation{Low Temperature Laboratory, O.V. Lounasmaa Laboratory, Aalto University, P.O.Box 15100, 00076 AALTO, Finland}
\author{Visa Vesterinen$^{\dagger}$}
\affiliation{VTT Technical Research Centre of Finland, P.O. Box 1000, 02044 VTT, Finland}
\author{Juha Hassel}
\affiliation{VTT Technical Research Centre of Finland, P.O. Box 1000, 02044 VTT, Finland}
\author{Heikki Sepp\"a}
\affiliation{VTT Technical Research Centre of Finland, P.O. Box 1000, 02044 VTT, Finland}
\author{Pertti Hakonen$^*$}
\affiliation{Low Temperature Laboratory, O.V. Lounasmaa Laboratory, Aalto University, P.O.Box 15100, 00076 AALTO, Finland}


\begin{abstract}
\vspace{2.0cm}
The fundamental noise limit of a phase-preserving amplifier at frequency $\omega /2\pi $ is the
standard quantum limit $T_{q}=\hbar \omega /2k_{B}$.
In the microwave range, the best candidates have been amplifiers based
on superconducting quantum interference devices (reaching the noise temperature $T_{n} \sim 1.8\mathbf{T_{q}}$ at 700
MHz), and non-degenerate parametric amplifiers (reaching noise levels
close to the quantum limit $T_{n}\approx \mathbf{T_{q}}$ at 8 GHz). We
 introduce a new type of an amplifier based on the negative
resistance of a selectively damped Josephson junction. Noise performance of our amplifier is limited by mixing of quantum noise from Josephson
oscillation regime down to the signal frequency. Measurements yield
nearly quantum-limited operation, $T_{n}=(3.2\pm 1.0)\mathbf{%
T_{q}}$ at 2.8 GHz, owing to self-organization of the working point.
Simulations describe the characteristics of our device well and
indicate potential for wide bandwidth operation.

\vspace{4.0cm}

$^{\dagger}$These authors contributed equally.

\end{abstract}
\maketitle

The goal of quantum limited amplification
at microwave frequencies has become increasingly important for superconducting qubits and nanoelectromechanical
systems \cite{R16}. The lowest noise temperatures with respect to the
quantum noise have been achieved using nondegenerate parametric
amplifiers based on superconducting quantum interference devices
(SQUIDs) \cite{R06, R07, R08}. They yield a noise
temperature $T_{n}$ of about $\left( 1.0-1.6\right) \mathbf{T_{q}}$.
Other implementations of near-quantum limited amplification have been
realized by means of Josephson ring oscillators \cite{R09}, DC-SQUIDs
\cite{muck,clarke}, and parametric amplifiers based on Josephson
junction arrays \cite{R01,R02,R04,R05,R06}. Devices based on
photon-assisted tunneling SIS-mixers yield
$T_{n}=1.2\mathbf{T_{q}}$ \cite{R10}. However, these devices lack power
gain but they do have a large gain in photon number due to conversion
from high to low frequency.

Negative differential resistance devices, in particular tunnel
diodes, have been used in the past to construct oscillators and
amplifiers for microwave
frequencies. These devices are capable of very fast operation. They were
among the first ones to be used at microwave frequencies because they
display little or
no excess noise in the negative resistance bias region \cite{R11}. Here,
we
propose a negative-resistance amplifier based on an unshunted, single
Josephson
junction (JJ) operating in a noise compression mode. Unshunted junctions
have been analyzed and demonstrated to work in SQUID circuits at low
frequencies by Sepp\"{a} et al. \cite{R12}. We have developed analogous
concepts for high frequency operation. The present device differs
markedly from previous implementations using unshunted Josephson devices
due to the modified impedance environment.

Unshunted junctions are attractive as low-noise devices since they minimize
fluctuations by avoiding unnecessary dissipation in the junction
environment. In voltage-biased ($V_\mathrm{b}$) operation, these devices can be considered as mixers
between the signal frequency ($\omega _{s}$ around a few GHz) and the
Josephson frequency ($\omega _{J} =\left( 2e/\hbar \right) V_\mathrm{b}=2\pi \times 10-300$ GHz)
including sidebands 
\cite{likharev}. A frequency-dependent environmental impedance can be employed
for controlling mixing strengths (because the Josephson junction is a
phase driven current generator) and the impedance makes the conversion
between these two quantities.

\section*{RESULTS}

The fundamental macroscopic principle of our single junction amplifier (SJA)
is that the intrinsic resistance of a JJ is negative over time scales much
longer than $1/ \omega_J$ \cite{R12} (as shown in Fig. 1a).
This is
usually hidden in weakly damped JJs since the negative-resistance branch is
unstable. On the other hand, for strongly damped junctions, the total
dynamic
resistance is positive. This can be seen from the current-voltage
$IV$
characteristics $v_\mathrm{b} =
\sqrt{i_\mathrm{b}^2-1}$ for a Josephson junction with negligible capacitance (valid for $i_\mathrm{b}>1$). Here $v_\mathrm{b} = V_\mathrm{b}/I_cR$ denotes the voltage scaled with critical current $I_c$ and the shunt resistance $R$ while $i_\mathrm{b} = I_\mathrm{b}/I_c$ is the dimensionless current. Solving for the current through the
junction alone, $i_\mathrm{JJ} = i_\mathrm{b}-v_\mathrm{b}$ (illustrated by the black curve in Fig. 1a), we get
for the scaled dynamic resistance
\begin{equation}
r_d=\frac{R_d}{R}=\frac{1}{di_\mathrm{JJ}/dv_\mathrm{b}} = \frac{1}{di_\mathrm{b}/dv_\mathrm{b}-1}
\label{rdyn}
\end{equation}.
This yields $\sqrt{v_\mathrm{b}^2+1}/(v_\mathrm{b}-
 \sqrt{v_\mathrm{b}^2+1})$, negative at all bias points.

\begin{figure}[!ht]
\begin{center}
\includegraphics[scale=0.47]{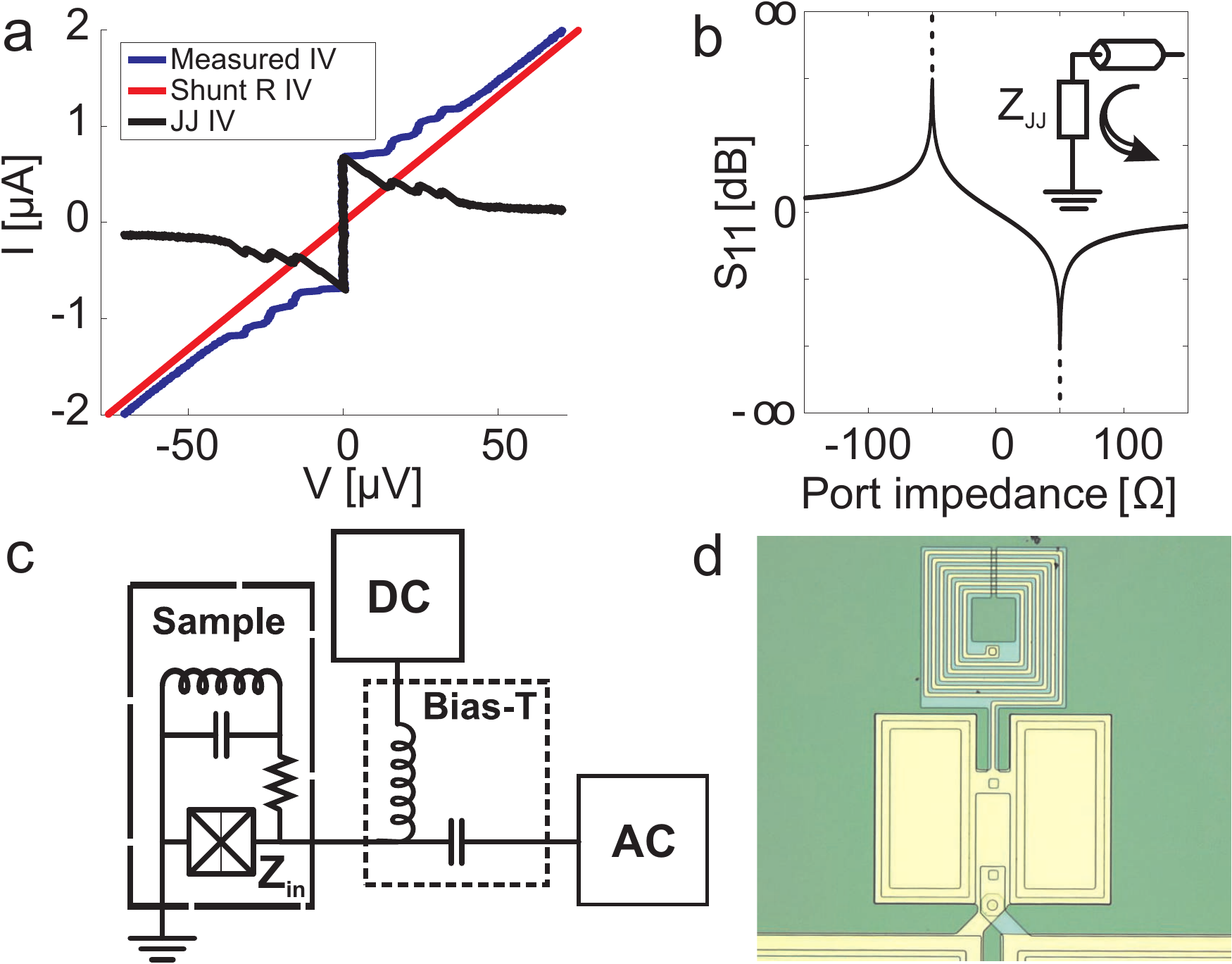}
\end{center}
\caption{ a) Typical IV of a SJA (in blue); red and black curves indicate
the division of $I$ into shunt and junction currents, respectively. b)
Reflection (scattering) amplitude $S_{11}$ in a $Z_0 = 50$ $\Omega$ system
as a function of the load impedance. c) Principal scheme of the SJA
operation.
d) Optical image of a SJA; the size of the image is approximately 270 $\protect\mu$m $\times$ 230 $\protect\mu$m.
}
\label{fig1}
\end{figure}

The schematics of our SJA configuration is illustrated in Fig. 1c.
To
utilise the negative resistance of a JJ for amplification, stable operation has
to be maintained by sufficient damping at all frequencies. The
frequency-dependent damping is set in such a way that the external shunt
damps the low ($\omega < \omega_s$, the signal frequency) and high ($\omega
>\omega_s$) frequency dynamics, which ensures both stable DC bias and overdamped Josephson dynamics. In practice, we have realised this
separation by mounting the shunt resistor in series with a
bandstop filter whose center frequency is at the signal frequency
$\omega_s$ \cite{note, note2}. The shunt capacitor is chosen large
enough
that it acts as a
short at the Josephson frequency to ensure the high frequency dynamics
and the IV curve are not modified. The
stabilization in the stop band is provided by the postamplification circuit.
The shunt circuit and the postamplification circuit together guarantee the
stability of the device by generating a wide-band resistive environment for
the JJ.
Operated as a
reflection amplifier, the power
gain $|S_{11}(\omega)|^2=|\Gamma(\omega)|^2$ is determined by the reflection
coefficient
\begin{equation}
\Gamma (\omega )=(Z_{in}(\omega )-Z_{0})/(Z_{in}(\omega )+Z_{0}),
\label{gamma}
\end{equation}%
where $Z_{in}(\omega )$ is the impedance of the JJ, the shunt and the series inductance; $Z_{0}$ is the impedance of the readout circuit. As seen from
the curve in Fig. 1b, there is gain ($S_{11}>0$ dB) at all values of
negative resistance and a strong divergence around $Z_{in}=-Z_{0}$. In the
stopband of the shunt circuit, the input impedance $Z_\mathrm{in}(\omega_s)$
consists of the JJ (and possibly of an LC impedance transformer): it is
real and
negative. For $|R_{d}| \gtrsim Z_{0}$, large gain with stable operation can be obtained. For operating conditions where $|R_{d}|\gg Z_{0}$
impedance transforming circuits are employed to change the
reference level impedance $Z_{0}$,\emph{\ e.g.} from 50 $\Omega$
typical for standard RF technology to a level of 1 k$\Omega $
which is a typical value of $|R_{d}|$ for small Josephson junctions
at high bias voltages.

The dynamics of SQUID circuits can be analyzed using a Langevin
type of differential equation for the phase variable $\varphi $ across the
Josephson junctions \cite{R13}. Good agreement of such Langevin analysis
with measured experimental results has been obtained in the past \cite{R14,
R15}. In the semiclassical approach, the generalized Nyquist noise formula
by Callen and Welton \cite{CW} with the frequency dependence $0.5\hbar \omega
\coth \left( \hbar \omega /2k_{B}T\right) $ is employed as the colored noise
source in the differential equation \cite{R14,levinson,brandt}. At the
Josephson frequency, the semiclassical noise power per unit bandwidth is so large ($\propto \hbar \omega_J \gg k_B T$) that, after downmixing, it will have observable effects on the phase dynamics at the signal frequency $\omega _{s}$. Since the noise at $\omega_{s}$ is cut off from the Josephson junction by the bandstop filter (see Fig. 1c), direct noise from the shunt is avoided and only the down-mixed noise 
is present in our device. The absence of direct noise ensures good noise
characteristics for our SJA and this feature is one of the basic
differences
when comparing SJAs with traditional microwave SQUID amplifiers.

\subsection*{Experimental}

\begin{figure}[!ht]
\begin{center}
\includegraphics[scale=0.4]{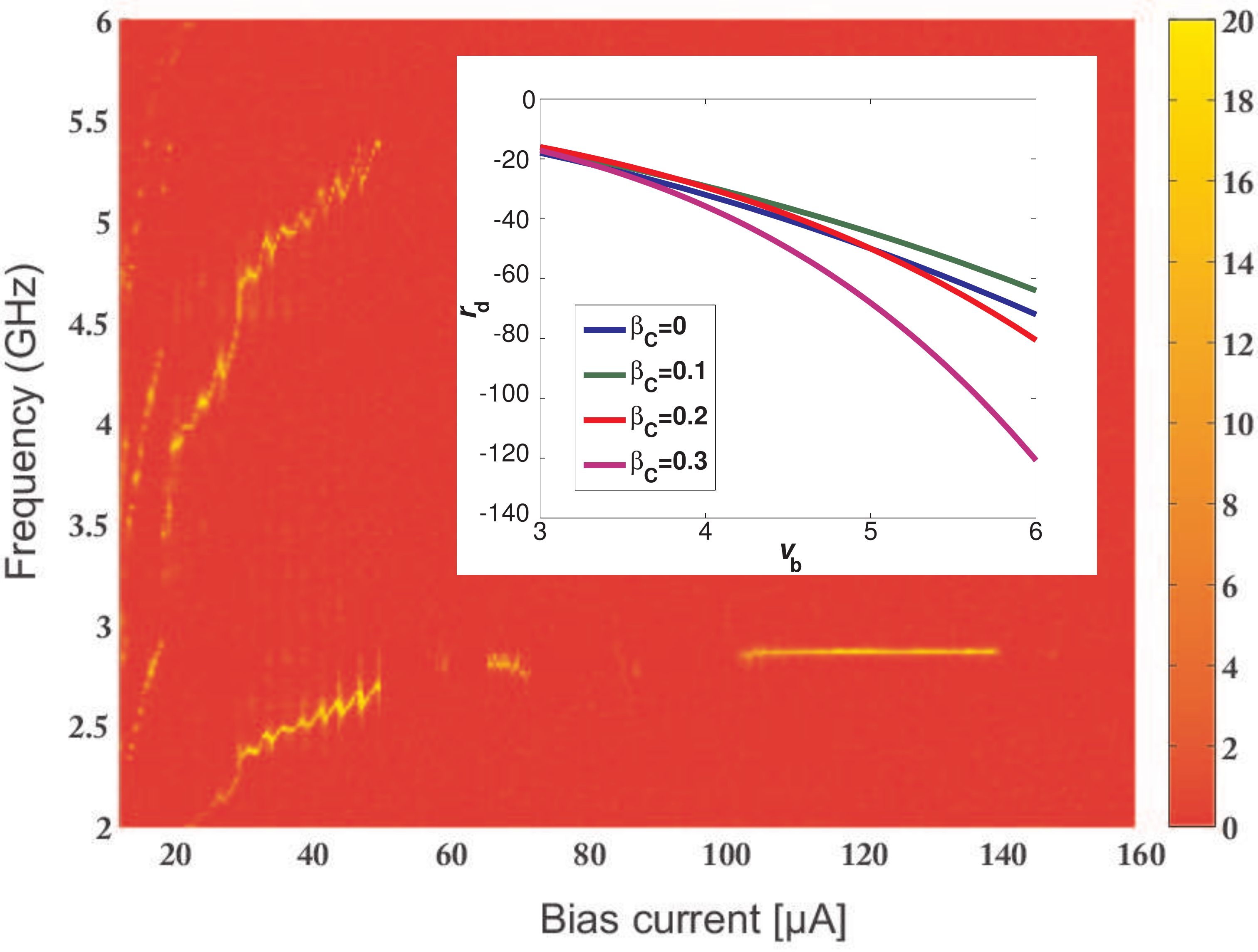}
\end{center}
\caption{Noise spectra of the device as function of the bias point; the
reference level corresponds to 14 K $k_B$ and the power scale on
the right is given in dB. The inset displays the dependence of $r_d$ as
a function of bias voltage for different values of $\beta_C$. Values
between 0 and $r_d^*$ lead to unstable behaviour; $r_d^*=-50$ without impedance transformer ($R=1\ \Omega$).
There are no special features in the noise spectral density in the area below the inset.}
\label{figyy}
\end{figure}

Fig. \ref{figyy} displays noise spectra measured on the device at different
bias points. At low bias currents, the magnitude of the dynamic
resistance $|R_d|$ is smaller than the environmental impedance in parallel to it, making the total damping
impedance of the LC resonator in the shunt circuit negative. This leads to either spontaneous oscillations or
saturation. The oscillations are highly nonlinear, which is manifested
as higher harmonics in the spectra. The saturation shows up as
vanishing
response. As $|R_d|$ increases at higher bias points, the system is
stabilized and the harmonics disappear since the device operates as a linear amplifier generating amplified
noise at the output.

After finding the optimal stable bias point,
the gain vs. frequency was recorded at several power levels. The maximum measured gain of the SJA was
found to be $28.3\pm 0.2$ dB.
The measured power gain of the device is plotted in Fig. \ref{figxx} at $P_\mathrm{in}=-160$ dBm.
The $-1$ dB
compression point for $P_\mathrm{in}$ was found to be around $-134$ dBm; this
yields a dynamic range of $70$ dB as the input noise corresponds to $-204$ dBm.
For the $-3$ dB bandwidth, we obtain $\mathrm{BW} \simeq 1$ MHz. However,
the bandwidth depends very much on the bias voltage due
to the variation of $R_{d}$ along the IV-curve, indicating that
fundamentally the device is capable of wideband gain. In the present experiments, we reached $|\Gamma |_{\mathrm{max}}\times
\mathrm{BW}= 40$ MHz for the voltage gain - bandwidth product.
The nominal parameters of the measured amplifier are given in
Table I in the Methods section.

The inset in Fig. \ref{figxx} displays the improvement of the signal to noise ratio when the SJA is switched on and operated at its maximum gain. Based on this improvement, we find that the input-referred noise power added by the amplifier is $220\pm 70$ mK ($0.5\hbar \omega /k_{B}\coth (\hbar
\omega /2k_{B}T)=90$ mK originating from the source has been
subtracted), which corresponds to $T_{n}\approx \left( 3.2\pm 1.0\right)
T_{q}$. The best noise temperature was obtained at the largest gain of the SJA.

\subsection*{Theoretical}

To theoretically model a single junction device with
arbitrary,
frequency-dependent environment with
$0<\beta_c=2eR^2(\omega)I_cC_J/\hbar<1$, we simulate numerically the electrical circuit on the basis of the DC and AC Josephson
relations which define a nonlinear circuit element having the properties:
$I_{\mathrm{J}}=I_{\mathrm{c}}\sin\varphi $ and
$V=\left( \hbar /2e\right) \partial \varphi /\partial t$. We have compared our numerical simulations with analytic methods
using an approximate model where we have adapted the resistively
and
capacitively shunted junction (RCSJ) approach to the modified environmental impedance of the SJA. Our
numerical and analytic models take into account the Callen and Welton
quantum noise from the environment semiclassically. Down-conversion
of the noise at $\omega_J$ is the
main quantity to be minimized for optimum performance.

\begin{figure}[!ht]
\begin{center}
\includegraphics[scale=0.47]{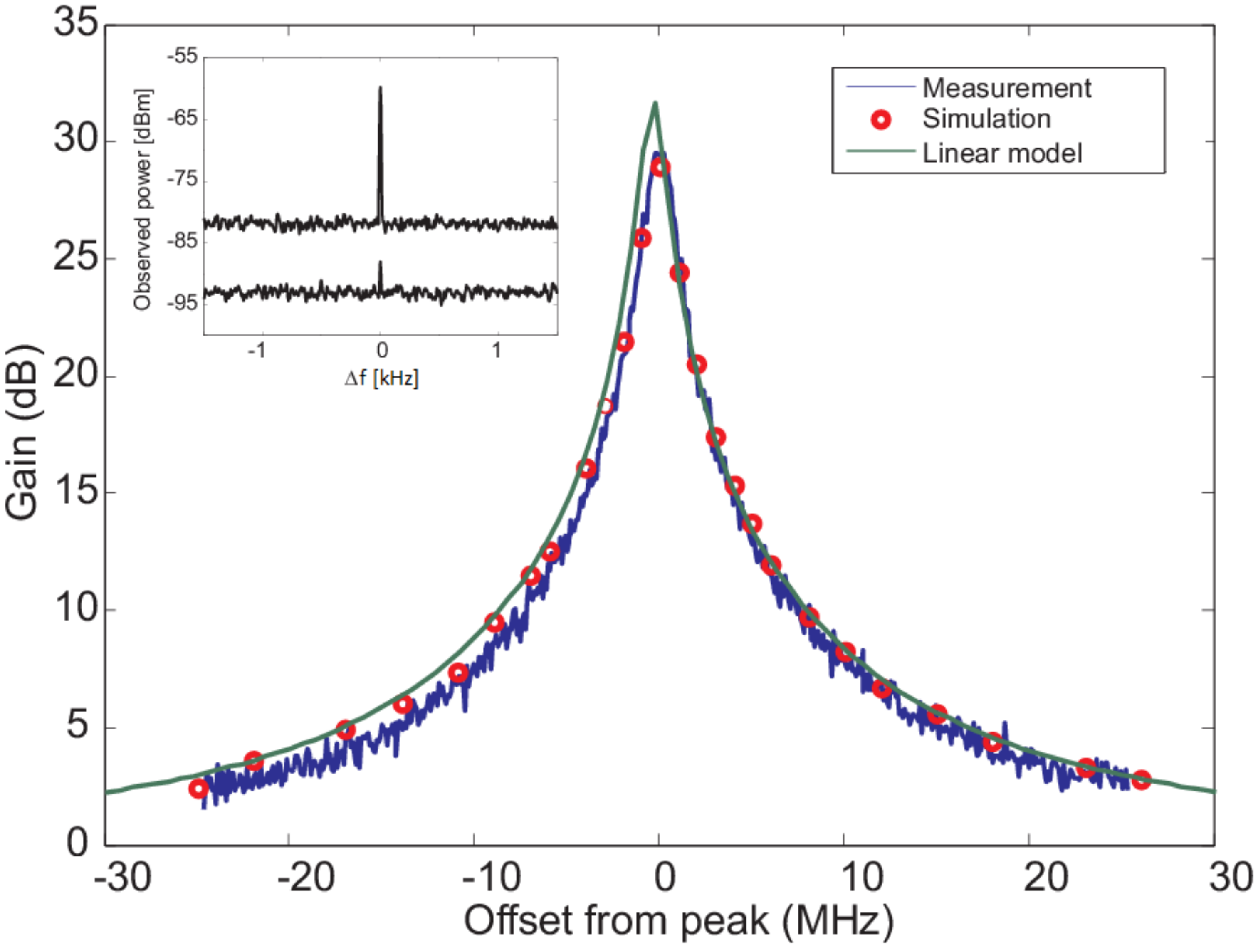}
\end{center}
\caption{Gain of the SJA as function of frequency at the optimal point of operation (blue, noisy curve). Results from our numerical simulation are denoted by open circles, while the smooth curve (green) illustrates the gain from a linearized electrical circuit model where the Josephson junction is
replaced by a negative resistance of $R_{d}=-1370$ $\Omega $ from Eq.~(\ref{rdyn}). Inset: Output noise spectra
 having the SJA off (lower trace) and on (at maximum gain).}
\label{figxx}
\end{figure}

The simulated power gain is included in Fig. \ref{figxx} together with
the experimental data. The theoretical gain curve is seen to follow the experimental behavior closely and it yields 42 MHz for the gain-bandwidth product. The simulated maximum gain amounts to $28.9\pm0.5$ dB. All these findings
are in excellent agreement with the experimental data. Basically, the
shape of the gain curve indicates that the amplification mechanism is
based on mixing between $\omega_s$ and the sidebands of $\omega_J$.
This occurs along with the conversion from down-mixed currents at
$\omega_s$ to
voltage by the shunt impedance (see the Supplementary material). For comparison, we have also calculated a linearized response curve where the Josephson junction has been replaced by a negative resistance of $R_{d}=-1370$ $\Omega $ from Eq.~(\ref{rdyn}).

Our numerical simulations yield $T_{n}=270\pm 30$ mK which is close to
the experimentally found $T_n=220\pm70$ mK. Hot-electron
effects were taken into account by using the model of Ref. \onlinecite{wellstood}, on the basis of which
we estimated the electronic temperature in the shunt to be $T_e \simeq
400$ mK instead of the base temperature 70 mK. The noise temperature is
not very sensitive to hot electron effects when the shunt is fully
blocked by the $LC$ resonator at the center frequency. However, when
going away from the center frequency, direct noise may leak out from the
shunt reducing the useful band to "a noise-temperature-limited" range.
The simulated noise power spectrum and the corresponding $T_n$ as a
function of frequency are presented in Fig. \ref{noisetemp}.

\begin{figure}[!ht]
\begin{center}
\includegraphics[scale=0.7]{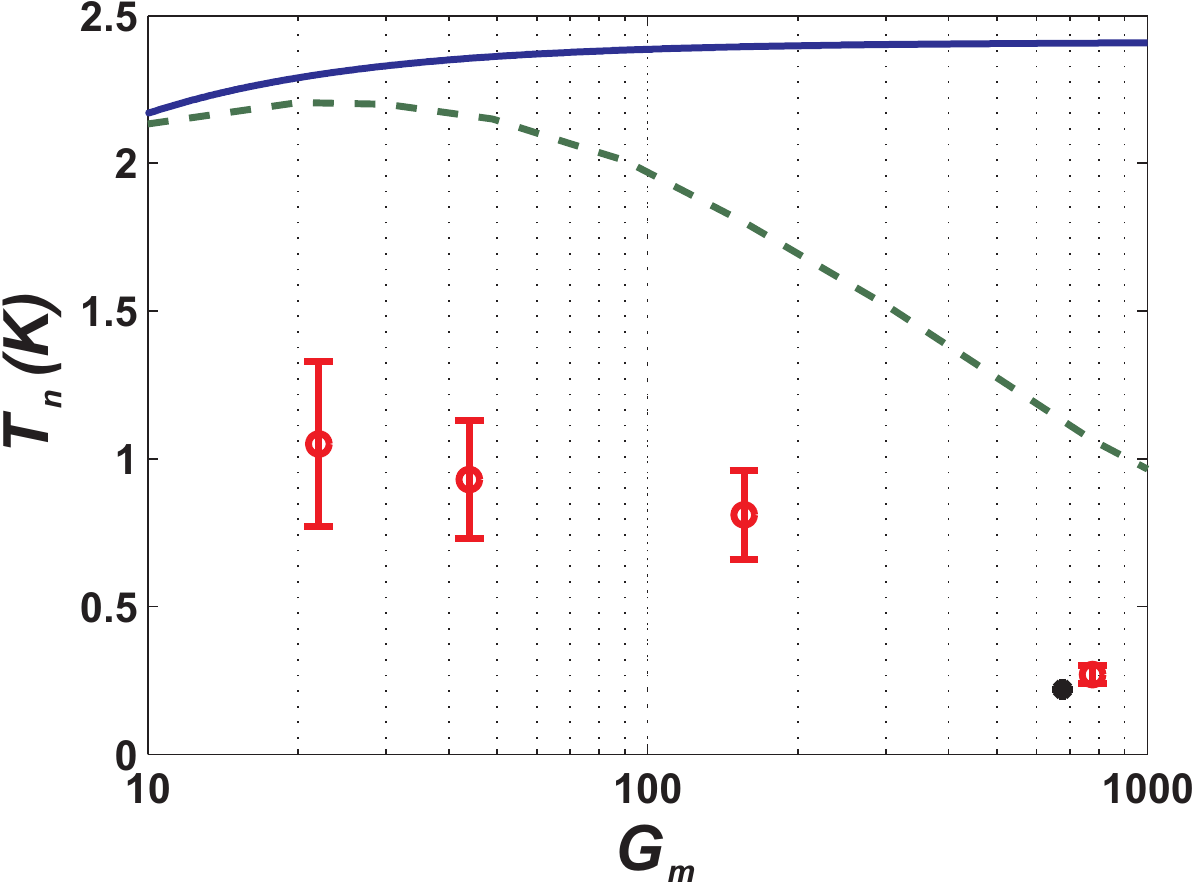}
\end{center}
\caption{Input noise temperature $T_n$ vs. maximum gain of
the SJA. The uncompressed $T_n$ (Eq.~(\ref{mixnoise})
with $\xi=1$, solid blue line) converges to $2.4$ K
at high gain. Compression suppressed $T_n$ for the analytic model with
two
sidebands at $\omega_J\pm \omega_s$ is denoted by
the dashed green line.
Noise temperature from the simulations is depicted using open circles, and the error bars represent the statistical
uncertainty in the simulated spectral density. The measurement result is
marked by a filled circle ($\bullet$), while the standard
quantum limit would be $T_q=\hbar \omega /2k_{B}=70$ mK.}
\label{noisegain}
\end{figure}


In our analytic modeling, we have generalized the semiclassical treatment of Ref.~\onlinecite{R14} to finite capacitance $C_J$ and combined the mixing analysis with the current-voltage characteristics derived in Ref. \onlinecite{brandt}. For the noise analysis, we define a noise process $\varphi_s(t)$, band-limited near the signal frequency.
Another noise process $\varphi_j(t)$ with $\langle \varphi_j(t)^2\rangle \ll 1$ covers the Josephson
frequency and one pair of sidebands ($\omega_J\pm \omega_s$). $\varphi_j$ has a small variance because of the low impedance of the junction capacitor at high Josephson frequency. We expand $i_J=\sin \varphi(t) \approx \sin (\omega_Jt+\varphi_s+\varphi_j)$
in order to describe the junction as a DC current generator plus two AC
current noise generators: one at $\omega_s$ and the other around the
Josephson frequency. In the Fourier plane, the AC Josephson relation and
the impedance environments at low and high frequencies establish the
down-mixing noise process. We denote the variance of the phase noise
over the signal band by $\delta_s^2=\langle \varphi_s(t)^2\rangle$. In
our calculations, we expand $\exp(i\varphi_s)\approx
J_0(r)+i(2J_1(r)/r)\varphi_s$ (which is a good approximation at small
$\delta_s^2$) but this breaks down when additional sidebands
($\omega_J\pm 2\omega_s$ and so on) become significant. These Bessel
functions of the first kind have the phase noise amplitude $r$
divided by the signal band. Ideally, $r$ should follow the Rayleigh distribution. In our analysis, we treat separately the limit of small fluctuations, $\delta_s^2 \ll 1$, and the regime with $\delta_s^2 \geq 1$, in which noise compression effects appear. With large gain and resonantly boosted current-voltage conversion, the phase fluctuations will grow so much that the non-linearities begin to limit the gain, and the system is driven to a steady state where the down-mixing process becomes altered and significantly suppressed. The number of added quanta per unit band from mixed-down noise is derived in the Supplementary material:
\begin{equation}
\frac{k_BT_\mathrm{mix}}{\hbar \omega_s}=\frac{N\xi(\delta_s^2)}{2}\left(
\frac{1+\beta_c^2v_\mathrm{b}^2}{1+3\beta_c^2v_{b}^2}\right)\left( \frac{G_m-1}{G_m}\right) ,
\label{mixnoise}
\end{equation}
where $N=\omega_J / \omega_s$ and the factor $(G_m-1)/G_m$ can be neglected at large gain. Noise suppression
is denoted by the compression factor $\xi(\delta_s^2)\leq 1$ which equals unity at $\delta_s^2 \ll 1$ and
decreases towards zero
with growing variance. In our model with the sidebands $\omega_J\pm\omega_s$, we obtain
$\xi(\delta_s^2)=<J_0^2(r)> \sim exp(-\delta_s^2)$. Hence, large
improvement in noise performance can be achieved compared to the
linear where $\xi(\delta_s^2)=1$.

The role of noise compression in the operation of the SJA is illustrated
in Fig. \ref{noisetemp}. For reference, we plot the
uncompressed noise from Eq.~(\ref{mixnoise}) multiplied by the simulated gain. The output noise
temperature from the actual simulation differs from it (an indication of
noise compression). The simulated
spectrum is rounded near the gain peak, which creates a dip in the input noise temperature.

\begin{figure}[!ht]
\begin{center}
\includegraphics[scale=0.6]{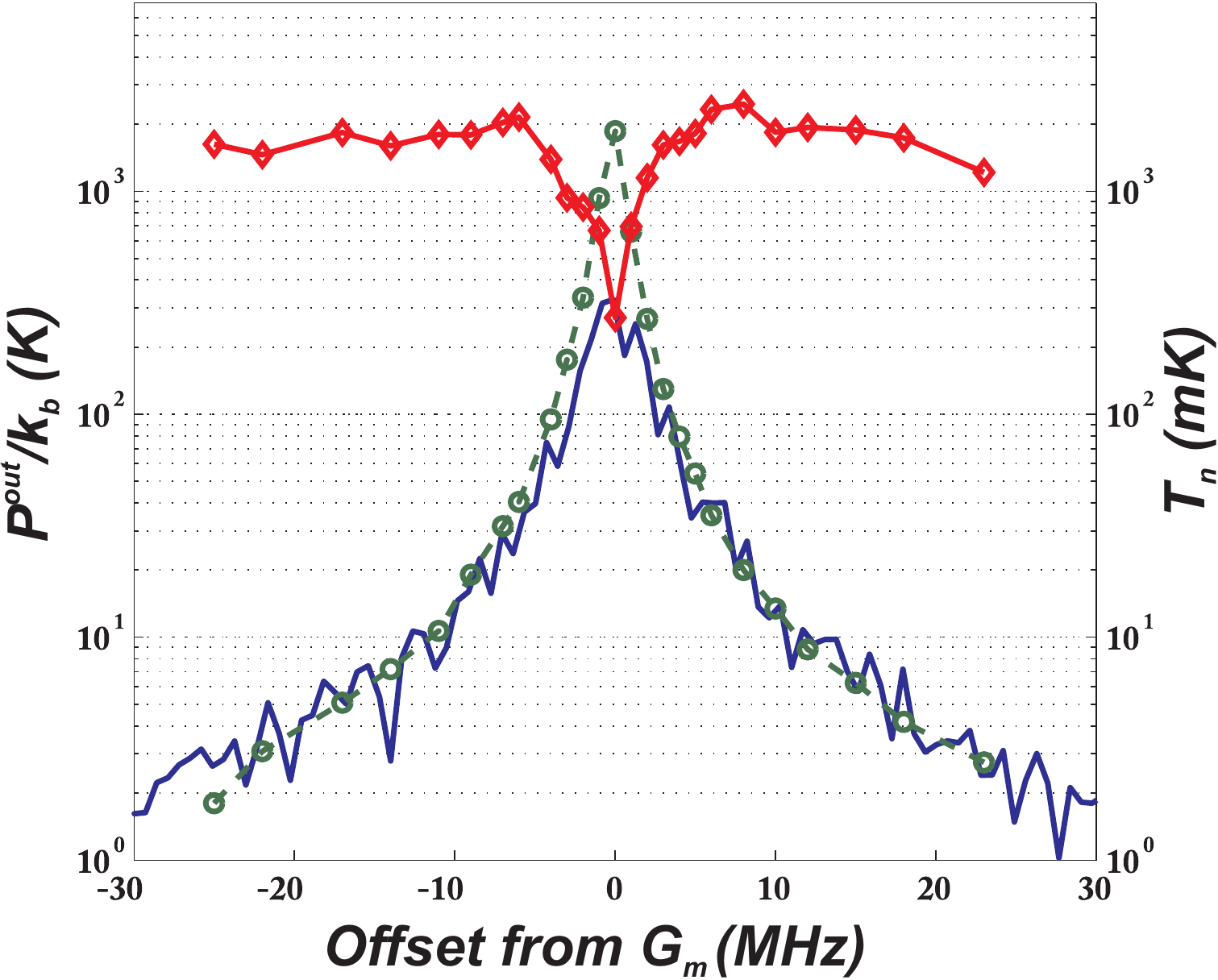}
\end{center}
\caption{The effective output noise temperature $P_n^{out}/k_B$ (left scale) is
compressed in the simulation (denoted by solid blue line) when compared with
the product of the simulated gain $(G-1)$ and the uncompressed
down-mixed noise of 2.4 K$ \times k_b$ from Eq.~\ref{mixnoise}
($\circ$). After dividing the simulated
output spectrum by the calculated gain, a clear dip is revealed in the input noise temperature ($\diamond$, right scale).}
\label{noisetemp}
\end{figure}

In Fig. \ref{noisegain}, the input
noise temperature at $G_{m}$ is plotted as a function of the gain.
Linear theories predict convergence towards $T_{n}=2.4$ K at $G_{m} >>
1$ (from Eq.~(\ref{mixnoise}) by taking $\xi=1$). Above a threshold gain
of $\sim 13$ dB, noise suppression sets in. From our analytic model with two sidebands $\omega_J\pm\omega_s$,
we obtain  $\xi =<J_0^2(r)>=0.44$ for the compression factor at $G_m=28$
dB and the noise temperature reduces to $T_{n}=1.0$ K. Compared with numerical simulations, the analytic model yields nearly $3-4$ times larger value for $T_n$.

\section*{DISCUSSION}

The compression mechanism for noise is crucial for the high bias operation of the SJA since otherwise $T_n$ would grow directly proportional to $v_\mathrm{b}$ ($N$ in Eq. (3)). The operation with noise compression can be
viewed as self-organization of the
system. Microscopic degrees of freedom give rise to a macroscopic order
which can be parametrized to describe the behavior of the system. In our device, the macroscopic ordering is dictated by the integrated noise over the amplified
bandwidth. This parameter governs the macroscopic characteristics of the
device (e.g. the effective critical current and the gain of the
device for external signals). The actual value of the
gain is set by the higher order terms present in the Josephson energy, which
resembles that of the order parameter stabilization in regular phase
transitions.

The bandwidth of our SJA is fundamentally limited below the Josephson and plasma
frequencies, $\frac{1}{2}\min (\omega _{J},\omega _{p})$.
It can be shown that the gain-bandwidth product is $|\Gamma
|_{\mathrm{max}}\times
\mathrm{BW}=2/|R_{d}|(C+C_J)$ in our first-order filtering scheme.
In the measured amplifier, the capacitance of the bandstop filter is
$C\approx 4.3 $ pF and $C_J=0.35$ pF.
Furthermore, using $R_{d}= -1370\ \Omega $
as in our operating point of interest, the formula yields $|\Gamma |_{\mathrm{max}}\times
\mathrm{BW}= 50$ MHz while $\approx  40$ MHz is obtained
experimentally.
In general, stability of the amplifier requires that $C>C_{J}$. Reduction of the shunt capacitance facilitates improvement of the gain-bandwidth product but the boundary condition $R\gg (\omega_JC)^{-1}$ must be met.
High bandwidth is predicted at small $R_d$ too, which can be obtained
most effectively by increasing the critical
current. Also $C_J$ controls the value of $R_d$ so that the
optimum for gain-bandwidth product is obtained for a small
junction with a high
critical current density.

Another possible low noise regime for the SJA is the limit of
small $\omega_J$. We analyzed a few devices at
$v_b=3$ ($N=2.33)$ with different $\beta_c$ (see the Suppl.). We obtained analytically
that the down-mixed noise contribution is around $\hbar \omega$ at $\beta_c =0.3-0.5$ without any noise
compression. This was verified in numerical simulations
according to which $0.9\pm 0.2$ quanta
were added by our SJA. Addition of one quantum indicates that the noise behaviour of the SJA is reminiscent
to that of heterodyne detection where the image frequency brings an extra noise of $\frac{1}{2}\hbar\omega$
to the detected signal \cite{Haus}, i.e. both sidebands of the Josephson frequency add $\frac{1}{2}\hbar\omega$ to the
noise temperature.

\begin{figure}[!ht]
\begin{center}
\includegraphics[scale=1.33]{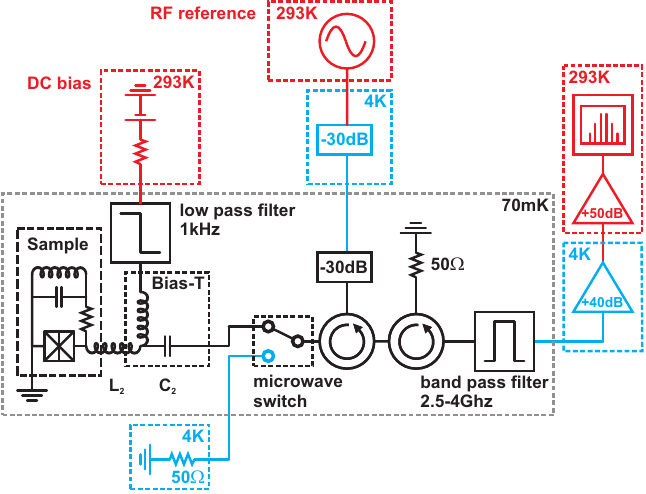}
\end{center}
\caption{ Setup for measuring the SJA characteristics. The essential components of the SJA
are located at 70 mK (indicated by the dashed black box). 60 dB of
attenuation is employed to thermalize the incoming rf signal cable and two
circulators eliminate the back action noise from the preamplifier. Noise
temperature of the cooled preamplifier (including losses in front of the
preamplifier)
$T_{n}^\mathrm{HEMT}=14\pm 3$
K at the center frequency of the SJA.}
\label{fig2}
\end{figure}

\begin{table}[!ht]
\centering
\caption{SJA parameters in the experiment and the simulation. Definitions: $Z_0$, impedance of the source
and the readout circuit; $R$, $C$ and $L$ the shunt resistance, capacitance and inductance, respectively;
$I_c$, $C_J$, $\omega_p$ and $\beta_c$ the critical current, the capacitance, plasma frequency and the
Stewart-McCumber parameter of the junction, respectively; $C_2$, $L_2$ the capacitance and the
 inductance in series with the SJA device (impedance transformer); $\omega_s$ the signal frequency; $I_b$ and $\omega_J$ the bias current
and the Josephson frequency at the optimal operating point.}
\label{tab:params}
\vspace{1.5cm}
\begin{tabular}{|ll|ll|}
\hline
Parameter & Value & Parameter & Value \\
$Z_0$ & 50 $\Omega$ & $I_\mathrm{c}$ & 17 $\mu$A\\
$R$ & 4.0 $\Omega$ & $C_\mathrm{J}$ & 0.35 pF\\
$C$ & 4.26 pF & $\omega_\mathrm{p}/(2\pi)$ & 61 GHz\\
$L$ & 702 pH & $\beta_c$ & 0.29\\
\hline
$C_2$ & 33 pF & $I_\mathrm{b}$& 140 $\mu$A\\
$L_2$ & 14.25 nH &$\omega_\mathrm{J}/(2\pi)$&270 GHz\\
$\omega_s$ & 2.865 GHz  & & \\
\hline
\end{tabular}
\end{table}

\clearpage

The control of noise in our SJA is not fully optimized and several issues
should be addressed in order to make the theoretical procedure for noise
minimization more effective and transparent. Using numerical simulations, we
 reproduced the measured noise temperature $3.2T_{q}$ at high bias and
found signs for the complex behavior of our device. Our
analytical model mixes down noise only from two sidebands
$\omega_J\pm\omega_s$, the consideration of which
is sufficient at low Josephson frequency
and small phase noise variance $\delta_s^2$. Consequently, the
predictions of $T_n \sim \hbar\omega$ from our analytical modeling
are reliable at low bias voltage. In the noise compression mode, $\delta_s^2 \geq
1$, our simulations show that the analytic model fails and an extension in the number of tracked sidebands is necessary.
 Moreover, further work will be needed to show whether pronounced noise
compression can drive the SJA into the standard quantum limit $T_q$. Our
analysis indicates that the concept of selectively shunted
junction
amplifier for microwaves is sound and that it provides the best route for quantum limited operation over large bandwidths.

\section*{METHODS}

Our experimental setup for the SJA measurements is shown in Fig. \ref{fig2}. The
device is biased with a DC current which allows the effective value of the
negative resistance to be tuned over a wide range of values. The incoming signal and the reflected signal
are separated by circulators and the signal postamplification is performed by high electron mobility
transistor (HEMT) based amplifiers at 4 K and at the room temperature.
At the optimal operating
point, the dynamic resistance $R_d$ of the Josephson junction is -1370
$\Omega$ in our amplifier. To get
substantial gain according to Eq. 2, we apply impedance transformation
by placing an inductor $L_2$ in series
with the junction. This converts the input impedance $Z_{in}(\omega_s)$
close to -50$ \Omega$.


To measure the amplifier performance, we injected a reference signal and recorded the signal-to-noise (S/N) ratio while having the
SJA ON and OFF. In the OFF state, the SJA acts like a pure inductance
reflecting all the incoming power (passive mirror) and the noise in the S/N
ratio measurement is fully specified by the HEMT preamplifier. The largest improvement
in the S/N was found at the highest bias current $\sim 140$ $\mu $A ($i_\mathrm{b}=8.2$). Using a
source at 70 mK, the S/N ratio after the HEMT amplifier was improved by $%
17.2\pm 0.2$ dB. Thanks to the microwave switch in the setup, the noise temperature of the HEMT amplifier could be carefully calibrated using the cold/hot load technique.
The parameters of the investigated amplifier are
collected into Table I.

\section*{ADDITIONAL INFORMATION}

\subsection*{Acknowledgements}
We thank Jari Penttil\"{a} for providing the Josephson junctions used in our
experiments. We also thank Thomas Aref for his support during the
writing of this article. This work was
supported by Center of Excellence and Materials
World Network grants of the Academy of Finland and by the European Science
Foundation (ESF) under the EUROCORES Programme EuroGRAPHENE.

\subsection*{Author Contributions}
All authors took jointly part in the planning of this experimental work
and the development of its theoretical interpretation. P.
L. performed the experiments and V. V. made the
numerical simulations and the analytical mixing analysis. P. L. and V.
V. co-operatively wrote the first versions of the main manuscript and
the Supplementary material. All authors
contributed to editing of the manuscript.

\subsection*{Competing Financial Interests}
The authors declare no competing financial interests.

\end{document}